\documentclass[structabstract]{aa}  
\usepackage{graphicx}
\usepackage[dvips]{color}
\usepackage{txfonts}
\usepackage{natbib}
\bibpunct{(}{)}{,}{a}{,}{,}

\begin{document}

   \title{A method to determine distances to molecular clouds using near-IR photometry}
   \titlerunning{}

   \author{
          Maheswar, G.
          \inst{1}\fnmsep\inst{2}
          ,
          Lee, Chang Won\inst{1}
          ,
          Bhatt, H. C.\inst{3}
          ,
          Mallik, Sushma V.\inst{3}
          ,
          Sami Dib\inst{4,5}
          }

   \institute{
Korea Astronomy and Space Science Institute, 61-1, Hwaam-dong, Yuseong-gu, Daejeon 305-348, Republic of Korea\\
              \email{maheswar@kasi.re.kr, maheswar@aries.res.in}
\and
   Aryabhatta Research Institute of Observational Sciences, Manora Peak, Nainital 263 129, India
\and
   Indian Institute of Astrophysics, Koramangala, Bangalore 560 034, India
\and
   Service d'Astrophysique, DSM/Irfu, CEA/Saclay, F-91191 Gif-sur-Yvette Cedex, France
\and
   Lebanese University, Faculty of Sciences, Department of Physics, El-Hadath, Beirut, Lebanon
             }

   \date{Received --- / Accepted ---}
 
  \abstract
  {}
   {To develop a method to determine distances to molecular clouds using $JHK$ near-infrared photometry.}
   {The method is based on a technique that aids spectral classification of stars lying towards the fields containing the clouds into main sequence and giants. In this technique the observed ($J-H$) and ($H-K_{s}$) colours are dereddened simultaneously using trial values of $A_{V}$  and a normal interstellar extinction law.  The best fit of the dereddened colours to the intrinsic  colours giving a minimum value of $\chi^{2}$ then yields the corresponding spectral type and $A_{V}$ for the star. The main sequence stars, thus classified, are then utilized in an $A_{V}$ versus distance plot to bracket the cloud distances.}
   {We applied the method to four clouds, L1517, Chamaeleon I, Lupus 3 and NGC 7023 and estimated their distances as $167\pm30$, $151\pm28$,  $157\pm29$ and $408\pm76$ pc respectively, which are in good agreement with the previous distance estimations available in the literature.}
   {}

  \maketitle


\section{Introduction}
The knowledge of distance to interstellar clouds which are either isolated or associated with large complexes, star forming or starless, is vitally important to provide observational constraints  to the properties of clouds and pre-main sequence (PMS) stars (e.g., Yun \& Clemens 1990; Clemens, Yun \& Heyer 1991; Kauffmann et al. 2008). However, distances to many of these clouds, especially those which are isolated, are highly uncertain (in some cases, by a factor of 2, see Hilton \& Lahulla, 1995).
 
In recent years, efforts have been made successfully to estimate distances to large  complexes using a number of methods. For example, combining  extinction maps from the Two Micron All Sky Survey (2MASS) with Hipparcos and Tycho parallaxes, Lombardi et al. (2008) estimated accurate distances to the Ophiuchus and Lupus dark cloud complexes; the Hipparcos parallax measurements and B-band polarimetry were utilized by Alves \& Franco (2007) to estimate the distance of the Pipe nebula. Recently, using multi-epoch very long baseline array (VLBA) observations of a number of young stellar objects (YSOs), Loinard et al. (2007) measured the trigonometric parallaxes and hence distances to them with a precision of about 1-4\% (Loinard et al. 2007, 2008). The distances to the parent clouds which harbour these YSOs are also constrained with similar precision.

However, it is difficult to establish distances to isolated clouds devoid of low-mass YSOs and  primary indicators such as ionizing stars or reflection nebulae. Distances to these clouds can be assigned through their association (in position-velocity space)  with nearby larger molecular clouds (e.g., Kauffmann et al. 2008) which have their distances determined through the above mentioned methods.  The traditional ways of determining distances to isolated clouds utilized the star count method (Bok \& Bok 1941) or Wolf diagrams (Wolf 1923) which is limited due to its dependence on the questionable extrapolation of luminosity functions. Additional methods of assigning distances to small dark clouds entail bracketing the cloud distance by using spectroscopic distances  to stars (Hobbs et al. 1986) and from the equivalent widths of interstellar Ca II H and K lines of stars close in front of and behind the cloud (Megier et al. 2005).  However, spectroscopy of a large number of stars projected on to a cloud is extremely demanding in terms of telescope time. Because of the absence of nearby associated large molecular clouds or stellar spectroscopic data, distances of most of the isolated small clouds remain unknown.
    
Distances to clouds can be determined by bracketing them using spectral classification of stars projected on to the cloud purely by photometry carried out in the Vilnius photometric system  (Strai\v{z}ys, 1991; Strai\v{z}ys et al. 1992). f The distance and reddening for individual stars can also be obtained using the very well calibrated Stromgren $uvby-\beta$ intermediate-narrow- band photometric system. The physical parameters of stars thus obtained are utilized to bracket the cloud distances (e.g., Nielsen et al. 2000; Franco 2002). Another innovative method was developed by Peterson \& Clemens (1998) by identifying M dwarfs lying both in front of and behind the cloud. The reddening and spectral types of these M dwarfs were determined from photometry alone due to their conspicuous position in ($V-I$) versus ($B-V$) colour-colour (CC) diagram.   The optical $V$, $R$, $I$ and near-IR (NIR) $J$, $H$ \& $K$ 2MASS colours were utilized by  Maheswar et al. (2004, 2006) to bracket cloud distances, again, by spectral classification  of stars projected on to them from photometry alone. 

The enormous NIR data provided by the 2MASS (Kleinmann et al. 1994) were extensively used  by several authors, mainly, (a) to discriminate between normal stars and stars with significant NIR emission from circumstellar material (e.g., Itoh et al. 1996) and (b) to map the extinction in dense molecular clouds (Cambr\'{e}sy et al. 2002;  Lombardi et al. 2006).  Dutra et al. (2003) build an $A_{K}$ extinction map of a $10^{\circ}$ field towards the Galactic centre using 2MASS $J$ and $K_{s}$ magnitudes. They utilized the upper giant branch of colour-magnitude diagrams and their dereddened mean locus built from previous studies of bulge fields to estimate the extinction. In this work we present a method by which distances to molecular clouds are estimated using ($J-H$), ($H-K_{s}$) colour indices of the main sequence stars in the spectral  range from A0 to K7. The technique is based on a methodology that provides spectral classification of stars projected on to the fields containing the cloud into main sequence and giants by simultaneously dereddening the ($J-H$) and ($H-K_{s}$) observed colours using trial values  of $A_{V}$ and a normal interstellar extinction law. The main sequence stars are then utilized in an $A_{V}$ vs. $distance~(d)$ plot to bracket the cloud distances. In \S\ref{method} we describe the method for finding the spectral type, $A_{V}$, and finally distances to the molecular clouds using NIR data from the 2MASS. The uncertainties involved in the determination of the $A_{V}$ values and the distances from the method are also discussed in this section. In \S\ref{data} we describe the criteria used to extract the data from the 2MASS database. In \S\ref{ChLuNGC} we apply our technique  to four clouds, L1517, Chamaeleon I, Lupus  and NGC 7023. We discuss the  method in determining distances to dark molecular clouds in \S\ref{discuss}.  Finally, we conclude the paper by summarizing the results obtained in \S\ref{conclu}. 
 
\section{The Method}\label{method}

The photometric distance $d$ to a star is estimated using the distance equation
\begin{equation}
d (pc) = 10^{(V-M_{V}+5-A_{V})/5} =10^{(K-M_{K}+5-A_{K})/5}\label{eq1}
\end{equation}
where $V$, $M_{V}$, $A_{V}$, $K$, $M_{K}$, and $A_{K}$ are the apparent magnitude, absolute  magnitude and extinction in $V$ and $K$ filters respectively. Spectral type and luminosity class information are needed to determine both absolute magnitude of a star and  the extinction suffered by it due to interstellar dust in the line of sight. Spectral type  and luminosity class also determine various colours of a star. Therefore, in principle,  one could determine the spectral type and the luminosity of a star from its observed colours  provided the extinction is zero. In practice, observed colours are reddened due to  interstellar extinction which is wavelength-dependent. By assuming a value for  $A_{V}$   and an extinction law one could estimate the colour excesses and  correct the observed colours to find the intrinsic colours of the stars and hence their spectral types.

\subsection{The ($J-H$), ($H-K_{s}$) intrinsic colours}\label{JHKCC}

\begin{figure}
\resizebox{8.5cm}{8.5cm}{\includegraphics{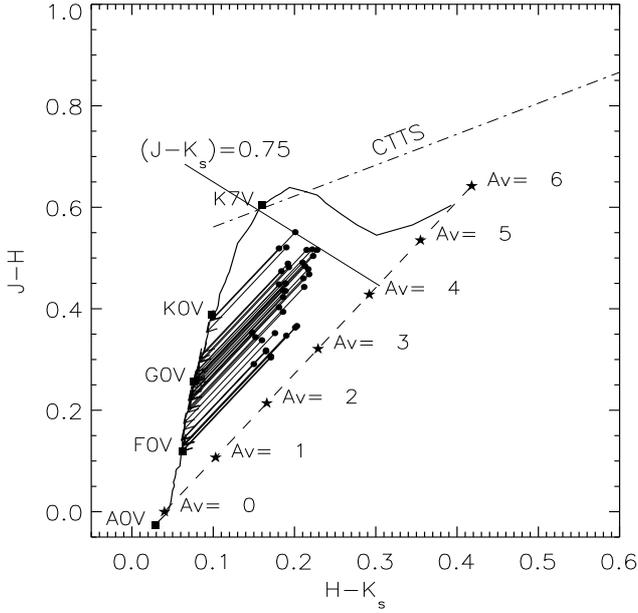}}
\caption{The ($J-H$) vs. ($H-K_{s}$) CC diagram drawn for stars towards a region chosen arbitrarily on the sky to illustrate the method described in \S\ref{method}. The solid curve represents locations of  unreddened main sequence stars. The reddening vector for an A0V type star drawn  parallel to the Rieke \& Lebofsky (1985) interstellar reddening vector is shown by the dashed line. The locations of the main sequence stars of different spectral types are marked  with square symbols. The region to the right of the reddening vector is known as the NIR  excess region and corresponds to the location of PMS sources.  The dash-dot-dash line represents the loci of unreddened CTTSs (Meyer et al. 1997). The closed circles represent the observed colours and the arrows are drawn from the observed to the final colours obtained by the method for each star.}\label{Fig1}
\end{figure}

In Fig.\ref{Fig1}, we present the ($J-H$), ($H-K_{s}$) CC diagram. The solid line represents locations of unreddened main sequence stars. Usually the unreddened colours of different spectral types and luminosity classes in the 2MASS system are  obtained from the Koornneef (1983) or Bessell \& Brett (1988) using the transformation equations given by Carpenter (2001). We produced the intrinsic colours of the main sequence stars in the spectral range from A0 to K7 in 2MASS system directly from the observations by selecting stars from the \textit{All-sky Compiled Catalogue of 2.5 million stars} (Kharchenko 2001) catalogue with known spectral types and which lie within 100 pc (estimated from their Hipparcos parallax  measurements) from us. The mean values of ($J-H$) and ($H-K_{s}$) colours produced  are listed in columns 3 and 4 of the Table \ref{tab_CC}, respectively.  The standard deviations and the number of stars used to evaluate the colours are given in columns 5, 6 and 7, respectively. The $M_{K}$ values obtained from  $M_{V}$ and ($V-K$) from Cox (2000) are listed in column 2. The intrinsic colors of the  main sequence stars and giants in the spectral range K8 to M6 were converted from Bessell \& Brett (1988)  to the 2MASS system using the transformation equations given by Carpenter (2001). We interpolated the intrinsic colour indices of normal main sequence stars and giants  to generate values with a 0.5 spectral sub-type interval.

\begin{table}
\centering
\caption{The estimated $(J-H)_{in}$ and $(H-K_{s})_{in}$ intrinsic colours for main sequence stars. 
These colours were estimated using 2MASS data of stars with known spectral types and
distances from the Hipparcos parallaxes. Only those stars located within 100 pc from the Sun were
chosen. The N$\star$ gives the number of stars used to estimate the colours.}\label{tab_CC}
\begin{tabular}{llccrrr}\hline
Sp.&$M_{K}$&$(J-H)_{in}$&$(H-K_{s})_{in}$&$\sigma_{(JH)_o}$&$\sigma_{(HK)_o}$&N$\star$\\
\hline  
A0V&   	0.65&	 -0.026&  0.029&0.030&0.028&50 \\
A1V&   	1.00&	 -0.014&  0.036&0.037&0.032&45 \\
A2V&   	1.17&	 -0.004&  0.043&0.044&0.035&43 \\
A3V&   	1.35&	  0.020&  0.048&0.049&0.032&58 \\
A4V&   	1.49&	  0.042&  0.050&0.044&0.026&23 \\
A5V&   	1.60&	  0.050&  0.052&0.040&0.029&24 \\
A6V&   	1.62&	  0.058&  0.051&0.037&0.024&12 \\
A7V&   	1.63&	  0.075&  0.054&0.039&0.031&23 \\
A8V&   	1.64&	  0.084&  0.055&0.047&0.032&8  \\
A9V&   	1.72&	  0.090&  0.059&0.033&0.026&37 \\
F0V&   	1.86&	  0.119&  0.062&0.042&0.026&105\\
F1V&   	2.00&	  0.142&  0.062&0.040&0.026&9  \\
F2V&   	2.15&	  0.162&  0.064&0.035&0.025&141\\
F3V&   	2.26&	  0.170&  0.065&0.032&0.029&269\\
F4V&   	2.38&	  0.175&  0.064&0.037&0.028&28 \\
F5V&   	2.49&	  0.195&  0.068&0.036&0.027&492\\
F6V&   	2.62&	  0.211&  0.072&0.036&0.027&390\\
F7V&   	2.75&	  0.228&  0.069&0.034&0.032&387\\
F8V&   	2.88&	  0.235&  0.069&0.038&0.029&308\\
F9V&   	3.04&	  0.252&  0.080&0.030&0.028&30 \\
G0V&   	3.19&	  0.257&  0.076&0.033&0.028&396\\
G1V&   	3.24& 	 0.265&  0.078&0.035&0.028&242\\
G2V&   	3.29& 	 0.276&  0.079&0.036&0.026&320\\
G3V&   	3.34& 	 0.285&  0.083&0.036&0.029&477\\
G4V&   	3.45& 	 0.302&  0.084&0.032&0.031&8  \\
G5V&   	3.56& 	 0.303&  0.085&0.035&0.030&499\\
G6V&   	3.68& 	 0.325&  0.090&0.040&0.028&188\\
G8V&   	3.90& 	 0.347&  0.093&0.048&0.030&259\\
G9V&   	4.03& 	 0.363&  0.094&0.074&0.021&9  \\
K0V&   	4.15& 	 0.388&  0.098&0.048&0.031&207\\
K1V&   	4.13& 	 0.406&  0.094&0.046&0.032&93 \\
K2V&   	4.15& 	 0.453&  0.115&0.053&0.028&140\\
K3V&   	4.15& 	 0.473&  0.120&0.046&0.039&137\\
K4V&   	4.22&  	 0.529&  0.130&0.057&0.037&43 \\
K5V&   	4.35&  	 0.555&  0.142&0.051&0.031&108\\
K7V&   	4.83&  	 0.604&  0.160&0.031&0.035&54 \\
\hline	
\end{tabular}
\end{table}

The locations of the stars with various spectral types are identified and marked in Fig.\ref{Fig1}. The dashed line is drawn parallel to the Rieke \& Lebofsky (1985) interstellar reddening vector. An unreddened main sequence normal star would move parallel to the reddening vector because of the extinction due to the interstellar dust along the line of sight. But as stars of spectral range  A0 to K7 move parallel to the reddening vector,  it would be difficult to differentiate between reddened stars in the spectral range A0 to K7 from unreddened M-type normal stars because the M-type stars lie across these vectors.  However, for the stars lying below M-type stars loci, it is certain that they are reddened or unreddened normal stars in the spectral range A0 to K7. The stars which occupy the region enclosed by the main sequence star loci (i.e., from A0-K7), reddening vector for an A0 type star and below M-type loci were utilized in this work to determine distances to the dark clouds.

\begin{figure}
\resizebox{9cm}{8.5cm}{\includegraphics{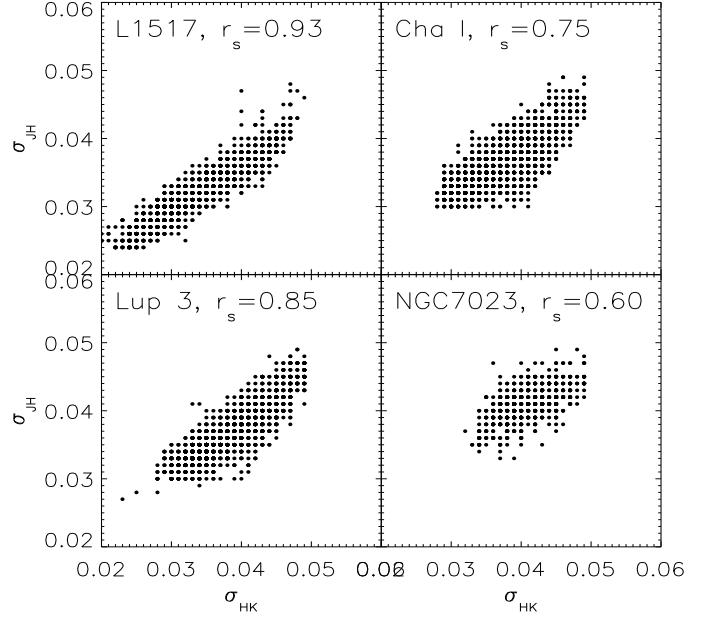}}
\caption{The plot shows the relation between the uncertainties in ($J-H$) and ($H-K_{s}$) colours of the sources in the four parts  of the sky studied in this work. The Spearman rank-order correlation coefficient, $r_{s}$, calculated for the regions are also given.  The $r_{s}$ values indicate a strong correlation between the uncertainties of the two colours.  The sources with the uncertainties  $\leq0.035$ magnitude in $J$, $H$, \& $K_{s}$ filters were considered.}\label{CC_Cor}
\end{figure}

\subsection{Spectral types \& $A_{V}$ from near-IR colours}

We used a procedure known as ``photometric quantification'', which is the procedure of inferring spectral type, and hence absolute magnitude and intrinsic colours of normal main sequence stars and giants from NIR photometry. In this method, stars with their ($J-K_{s}$)$\leq0.75$  
\footnote{This criterion allows us to avoid M-type stars from the analysis. Also, the classical T Tauri stars (CTTSs), often found to be associated the molecular clouds, occupy a well defined loci in NIR-CC diagram (Meyer et al. 1997) as shown in Fig. \ref{Fig1} which intercepts with the main sequence loci at ($J-K_{s}$)$\approx0.75$. Thus the criterion of ($J-K_{s}$)$\leq0.75$ would enable us to reject most of the CTTSs as well.} were chosen from the fields containing the cloud. Then a set of dereddened colours  [$(J-H)_{o}$ and $(H-K_{s})_{o}$] was produced for each star from their observed colours [($J-H$) and ($H-K_{s}$)] by using trial values of $A_{V}$ and the Rieke \& Lebofsky (1985) 
\footnote{The Rieke \& Lebofsky (1985) law was derived towards the Galactic center in the Arizona-Johnson, not 2MASS,  photometric system.   But the extinction law derived towards the Galactic center in 2MASS system by Nishiyama et al. (2009) though consistent with the transformed Rieke \& Lebofsky (1985) values, is different from those estimated by Indebetouw et al. (2005) derived towards massive star forming and ``field'' regions. It was suggested that the extinction law derived towards the Galactic center might be slightly different from those towards other sight lines. Since the maximum extinction that can be traced by our method is limited to $\sim4$ magnitude, the regions studied by us are in close resemblance to those by Indebetouw et al. (2005). However, the ratios, $A_{J}/A_{K}=2.51$ and $A_{H}/A_{K}=1.56$ from the Rieke \& Lebofsky (1985)  values are in good agreement with the average values of $A_{J}/A_{K}=2.5\pm0.15$ \& $A_{H}/A_{K}=1.55\pm0.08$  obtained by Indebetouw et al. (2005) suggesting that the untransformed values of the Rieke \& Lebofsky (1985)  do represent the reddening law in the 2MASS system. Similar suggestion was made by Alves et al. (1998) for their observations in CIT system. We used the values:  $A_{J}/A_{V}=0.282$, $A_{H}/A_{V}=0.175$, $A_{K_{s}}/A_{V}=0.112$, following  Cambr\'{e}sy et al. (2002) who found that the slope of the reddening vector measured in the 2MASS colour-colour diagram is in better agreement with the Rieke \& Lebofsky (1985) extinction law.} extinction law in the equations
\begin{equation}
(J-H)_{o}=(J-H) - 0.107\times A_{V},\label{eq2}
\end{equation}
\begin{equation}
(H-K_{s})_{o}=(H-K_{s}) - 0.063\times A_{V},\label{eq3}
\end{equation}
the trial values of $A_{V}$ were chosen in the range 0-10 magnitude with an interval of 0.01 magnitude though it is evident from  Fig. \ref{Fig1} that the maximum extinction that could be traced by this method is limited to $A_{V}\approx4$ magnitude. The computed set of dereddened colour indices for a star were then compared with the intrinsic colour indices [$(J-H)_{in}$ \& $(H-K_{s})_{in}$] of normal main sequence stars and giants, produced using the procedure discussed in the section \ref{JHKCC}. The best fit of the dereddened colours to the intrinsic  colours giving a minimum value of $\chi^{2}$ then yielded the corresponding spectral type and $A_{V}$ for the star. It is known that for a set of $N$ data points $y_{k}$ with associated uncorrelated uncertainties $\sigma_{k}$,  the best fit of predicted values $y_{k, mod}$ (from a model) to these points could be  obtained by $\chi^{2}$ minimization defined as (e.g., Gould 2003)
\begin{equation}
\chi^{2}=\displaystyle\sum_{k=1}^N\frac{(y_{k} - y_{k, mod})^2}{\sigma_{k}^{2}}\label{eq_chi}
\end{equation}
But while the uncertainties in the magnitudes on $J$, $H$ \& $K_{s}$ filters could be uncorrelated, the uncertainties in the two colours ($J-H$) \& ($H-K_{s}$) calculated from them with one filter ($H$) in common would be correlated.  In Fig. \ref{CC_Cor} we show the correlation between the uncertainties in colours of stars from the parts of the sky (L1517, Chamaeleon I, Lupus 3 \& NGC 7023) selected in this work. The Spearman rank-order  correlation coefficient, $r_{s}$, calculated for the sources in respective regions indicates a strong correlation between the uncertainties in the ($J-H$) and ($H-K_{s}$) colours. For correlated uncertainties, equation \ref{eq_chi} is not valid and one has to use (e.g., Gould 2003)
\begin{equation}
\chi^{2}=\displaystyle\sum_{k=1}^N\displaystyle\sum_{l=1}^N (y_{k} - y_{k, mod}){\bf B}_{kl}(y_{l} - y_{l, mod})\label{eq_chi1}
\end{equation}
where $\mathit B\equiv C^{-1}$, the inverse of the colour covariance matrix. In the case of the three bands considered here, the  colour covariance matrix is given as (Lombardi \& Alves 2001; Ma\'{i}z-Apell\'{a}niz  2004)
\begin{equation}
C_{J-H, H-K}=\left(\begin{array}{ccc} \sigma_{J}^{2}+\sigma_{H}^{2} & -\sigma_{H}^{2} \\ -\sigma_{H}^{2} &\sigma_{H}^{2}+\sigma_{K}^{2} \end{array}\right)
\end{equation}
with the off-diagonal elements having non-zero values unlike in the case of uncorrelated uncertainties. The expected value of $\chi^{2}$ after it has been minimized is given by the difference between the number of colours used and the number of parameters to be estimated (e.g., Gould 2003). Because we used two colours,  ($J-H$) and ($H-K_{s}$), to estimate two parameters, spectral type and extinction, the expected values of $\chi^{2}$ should be zero.  However, since we used discrete values of $A_{V}$ (0.01 magnitude) to deredden stars, the dereddened colours of the stars might not match exactly with the intrinsic colours of main sequence and giants to give  $\chi^{2}$ values  equal to zero but values $\ll1$. We considered only those solutions which gave a minimized $\chi^{2}\leq0.1$ for our analysis. 

The whole procedure is illustrated in the Fig.\ref{Fig1} where we plot the NIR-CC diagram for stars chosen from an arbitrary location on the sky. The arrows are drawn from the  observed data points to the corresponding dereddened colours estimated using our method. The maximum extinction values that can be measured using the method are those for A0V type stars ($\approx4$ magnitude).  The extinction traced by stars falls as we move towards more late type stars.

\subsection{Distances to dark molecular clouds}

Once the spectral type and  the $A_{V}$ of the stars are known, we calculate their distances using the equation
\begin{equation}
d~(pc) = 10^{(K_{s}-M_{K}+5-A_{K})/5}
\end{equation}
We excluded the stars classified as giants because of the large uncertainty in their absolute magnitudes. The distance at which the extinction of the stars showed a sudden increase in an $A_{V}$ vs. $d$  plot  was considered as the distance to the cloud. 

\subsection{Limitations and Uncertainties}

The uncertainty in the determination of distances to the clouds depends on the  uncertainty in the estimation of distances to the individual stars which is given by
\begin{equation}
\sigma_{d} =\sqrt[]{(\sigma_{K_{s}}^2+\sigma_{M_{K}}^2+\sigma_{A_{K}}^2)\times(d/2.17)^2}~~~~~~~~~(pc)\\\label{err_dist}
\end{equation}
where $\sigma_{K_{s}}$ is the photometric uncertainty in $K_{s}$ band, $\sigma_{M_{K}}$  is the uncertainty in the estimation of the absolute magnitude or the spectral types and  $\sigma_{A_{K}}$ is the uncertainty in the  $A_{K}$ ($=0.112\times A_{V}$) estimated by the method. The major sources of uncertainty are, however, the photometric errors and the uncertainty due to a different reddening law which are statistical and systematic in nature, respectively.

\begin{figure}
\resizebox{9cm}{16cm}{\includegraphics{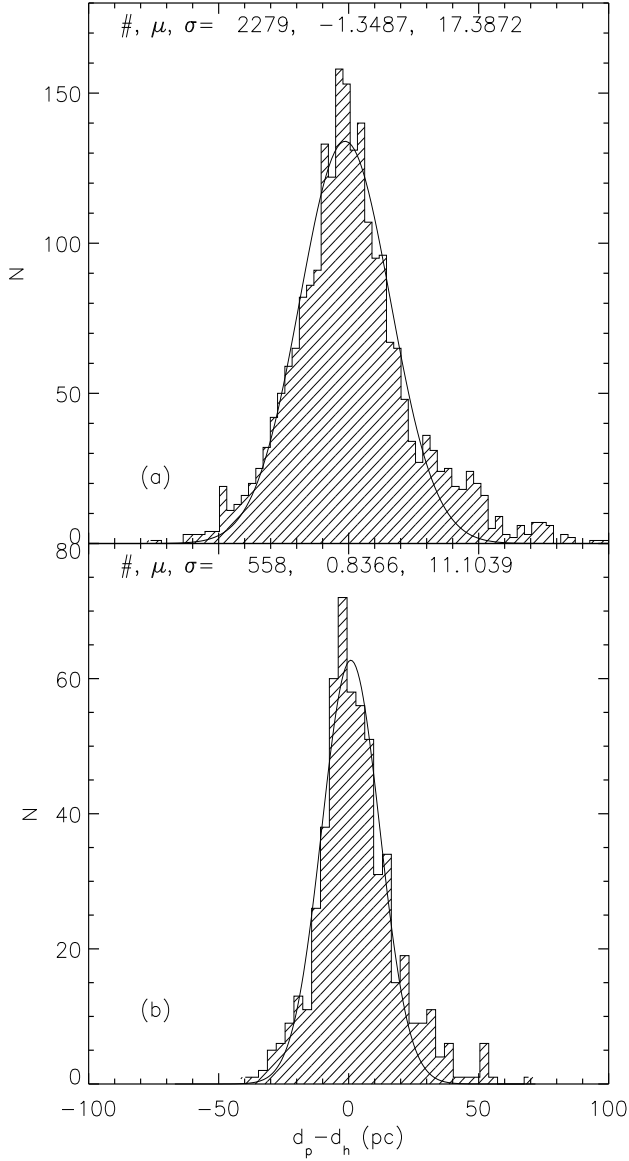}}
\caption{(a) Histogram of the difference between the distances of the main sequence stars from our method and  those from the Hipparcos parallax measurements with $\sigma_{\pi_{H}}/\pi_{H}\leq0.2$ are shown with a Gaussian fit  to the data over-plotted. (b) same as in (a) but for main sequence stars with $\sigma_{\pi_{H}}/\pi_{H}\leq0.05$.}
\label{hist1}
\end{figure}

We allowed a maximum photometric uncertainty \footnote{The photometric errors considered in this work  in the $J$, $H$, \& $K_{s}$ magnitudes from the 2MASS database include the corrected band photometric uncertainty, nightly photometric zero point uncertainty, and flat-fielding residual errors (Cutri et al. 2003).} of 0.035 magnitude for the sources in the  $J$, $H$, \& $K_{s}$ bands. The uncertainty in the absolute magnitudes or the spectral types  determined from our method was evaluated by applying it to stars with  known distances from the Hipparcos taken from the catalogue by  Kharchenko (2001). We used stars located within 100 pc from the Sun so that the effect of the  extinction could be neglected.  Then, the equation \ref{err_dist} applied to the Hipparcos stars becomes 
\begin{equation}
 \sigma_{d_{H}}=\sqrt[]{(\sigma_{K_{s}}^2+\sigma_{M_{K}}^2)\times(d/2.17)^2}  \label{err_dist1}
\end{equation}
In Fig. \ref{hist1}, we show histograms of the difference ($d_{p}-d_{h}$ pc) between the distances obtained using spectral types from our method, $d_{p}$ (pc) and those  from the Hipparcos parallax measurements, $d_{h}$ (pc). Figure \ref{hist1}(a) shows the distribution when stars with $\sigma_{\pi_{H}}/\pi_{H}\leq0.2$ (ratio of standard error to Hipparcos parallax) from Hipparcos parallax measurements were selected.  A Gaussian fit to the distribution shows a $1\sigma$ deviation of $\sim18$ pc. The scatter in the distribution reduced to $\sim11$ pc  (Fig. \ref{hist1}(b)) when stars with  $\sigma_{\pi_{H}}/\pi_{H}\leq0.05$  from Hipparcos parallax measurements were selected. The majority ($85\%$) of the latter stars were found   to be located at distances $\lesssim60$ pc. We noticed that the mean values of the distributions shown in both Fig. \ref{hist1}(a) and  Fig. \ref{hist1}(b) are shifted from the expected value of zero. Statistically the error in the mean, given by the $standard~deviation/\sqrt[]{N}$ ($N$ is the number of stars here), in both the distributions (a) and (b) should be less than 0.36 and 0.47 respectively. The negative shift in (a) could be due to the unaccounted binary stars which could  result in the underestimation of distances using photometry. Adopting a $1\sigma$ uncertainty of $18\%$ of the distance for the Hipparcos stars in the Eq. \ref{err_dist1}, the uncertainty in the $\sigma_{M_{K}}$ is estimated to be $\approx0.4$ which corresponds to an uncertainty in the spectral type of about four subclasses~\footnote{The change in $M_{K}$ with respect to the spectral types  in the range A0V-K7V could be fitted with a function, $M_{K}\approx0.1\times Sp.~type+0.98$.}.  We assumed 0.4 as the uncertainty in the $\sigma_{M_{K}}$ while calculating the uncertainty in the distances  for all the stars. 

The  $A_{V}$ values obtained for the stars from our method are those which when used to deredden the observed colours (in equations \ref{eq2} and \ref{eq3}) gave minimum values of   $\chi^{2}$.  Therefore the $A_{V}$ values obtained are equivalent to the average value of $A_{V}$ calculated using the individual colours, i.e., $A_{V}=[E(J-H)\times9.35+E(H-K_{s})\times15.87]/2$. Since the colours are found to be correlated, the expression for evaluating the uncertainty  in $A_{V}$ is
\begin{equation}
\sigma(A_{V}) =\sqrt[]{4.7^2.\sigma_{JH}^2+7.9^2.\sigma_{HK_{s}}^2+2\times37\times cov(JH, HK)}\label{err_Av}\\
\end{equation}
where, $\sigma_{JH}^2=\sigma_{J}^2+\sigma_{H}^2$, $\sigma_{HK_{s}}^2=\sigma_{H}^2+\sigma_{K_{s}}^2$
and $cov(JH,HK) = r_{s}\times \sigma_{JH}\sigma_{HK_{s}}$.
Using the maximum photometric uncertainty of 0.035 magnitude allowed in the individual bands and the highest value of 0.93 for $r_{s}$ (obtained towards L1517), the maximum uncertainty in the $A_{V}$ was found to be $\sim0.6$ magnitude. The uncertainty in the distances to the individual stars, equation \ref{err_dist}, is therefore dominated by  the uncertainty in the determination of the spectral types or $M_{K}$ values.

The true spectral type and the $A_{V}$ of the stars evaluated from our method directly depend on the reddening or extinction law that we adopt. Unlike in visual and UV wavelengths where the extinction law is known to vary along different sight lines in the Galaxy (Mathis 1990),  the similarity in the wavelength dependence of extinction for very different line of sights passing through different environments like the giant H II region (RCW 49) and the ``field'' region (Indebetouw et al. 2005) suggests that the extinction law is almost universal in the near-IR regime considered here. Consequently, the observed colours can be corrected for the reddening to determine the spectral type of the stars using a standard extinction law. But any deviation of the reddening law from the standard extinction law towards the region containing the clouds could introduce a systematic error in the distances estimated using our method.

\subsection{Determination of cloud distance from $A_{V}$-distance plot}\label{quant_approach}
\begin{figure}
\resizebox{9cm}{7cm}{\includegraphics{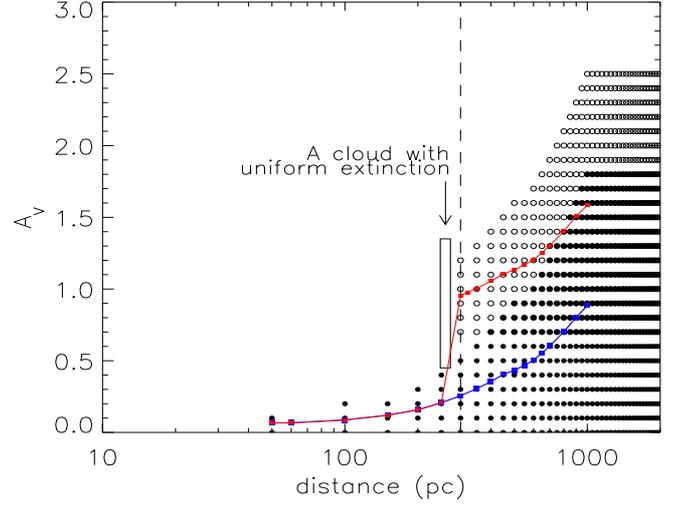}}
\caption{To illustrate the procedure used to determine distances to the clouds from an $A_{V}$ vs. $d$ plot as described in the \S\ref{quant_approach}.}
\label{cartoon}
\end{figure}
The presence of an interstellar cloud should reveal itself by a sudden onset of extinction at a certain distance. If the cloud is isolated and is the first interstellar feature encountered, the excess should almost equal to zero in front of the cloud (eg., Strai\v{z}ys et al. 1992; Whittet et al. 1997; Knude \& H{\o}g 1998;  Alves \& Franco 2006; Lombardi et al. 2008). In this technique, the distance to the cloud is  typically estimated from the first star that shows a significant reddening. The estimated distance is thus regarded as an upper limit due to the lack of information on the closeness of this background star to the cloud. In our method, because we used relatively large number of stars projected towards the cloud, we often found an almost continuous distribution of stars with high extinction beyond a certain  distance.

In Fig. \ref{cartoon} we show a schematic figure of an  $A_{V}$ vs. $d$ plot for an ideal situation along a line of sight through the Galactic plane. The extinction suffered by the stars increases as a function of their distance (typically, $\sim1.8~mag~kpc^{-1}$ is the value considered for the standard mean rate of  diffuse-ISM extinction in the Galactic plane near the Sun, Whittet 1992) represented by filled circles. If we group the stars in distance bins and calculate the mean $A_{V}$ of the stars for each bin, they  would follow the blue curve as shown in  Fig. \ref{cartoon}. The open circles represent the situation when a cloud of constant  extinction is introduced at a certain distance in the field-of-view. Now, the mean $A_{V}$ of the stars in the bins behind the cloud would show a jump equal to the extinction introduced by the cloud (red curve). The distance at which the first sudden increase in the  mean $A_{V}$ occurs could be considered as the distance to the cloud (marked by the vertical dashed line in  Fig. \ref{cartoon}).

In real cases, we grouped the stars into distance bins of $bin~width = 0.09\times distance$. The centers of each bin were separated by the half of the bin width. Because there exist very few stars at smaller distances, the mean of the distances and the $A_{V}$ of the stars in each bin were calculated by taking 1000 pc as the initial point and proceeding towards smaller distances (see Fig. \ref{4clds}). The mean distance of the stars in the bin at which a  significant drop in the mean of the extinction occurred was taken as the distance to the cloud and the average of the uncertainty in the distances of the stars in that bin was taken as the final uncertainty in our distance determination for the clouds.  For all the clouds studied in this work, the vertical dashed lines in $A_{V}$ vs. $d$ plots, used to mark the cloud distances,  were drawn at distances deduced from the above procedure (see  Fig. \ref{4clds}). The error in the mean values of $A_{V}$ were calculated using the expression, $standard~deviation/\sqrt[]{N}$, where $N$ is the number of stars in each bin.

\section{The Data}\label{data}

We extracted $J$, $H$ and $K_{s}$ magnitudes of stars from the 2MASS All-Sky Catalog of Point Sources (Cutri et al. 2003) which satisfied the following criteria,
\begin{enumerate}
\item Photometric uncertainty $\sigma\leq0.035$ in all the three filters.
\item Photometric quality flag of ``AAA'' in all the three filters, i.e., signal-to-noise ratio (SNR) $>10$.
\end{enumerate} 
The maximum allowed uncertainty of 0.035 magnitude in the photometry was chosen primarily as a trade-off between the minimum scatter in the data points and the availability of enough stars in the background and the foreground of the clouds to bracket the cloud distances. One added advantage of choosing the data with high quality was that the limiting magnitude becomes brighter and therefore, the selected stars  would sample the lower density regions of the clouds where a normal reddening law is believed to be the most valid (Strai\v{z}ys et al. 1982; Kandori et al. 2003).


\section{Distances to the molecular clouds L1517, Chamaeleon I, Lupus 3 and NGC 7023}\label{ChLuNGC}
 
We applied our method to four clouds, L1517, Chamaeleon I (Cha I), Lupus 3 (Lup 3) and NGC 7023, to determine distances to them. We preferred these four regions as they are structurally  less complex and have their distances estimated previously. Also, these clouds  harbour stars (inferred from associated nebulosity) for which distances from  the Hipparcos parallax measurements are available (van den Ancker et al. 1998;  Bertout et al. 1999) which would help constrain the parent cloud distances better. In Table \ref{dist_com}, we list the distances to them compiled from the literature.
\begin{table}
\centering
\caption{Distances to the clouds L1517, Cha I, Lup 3 and NGC 7023 compiled from the literature. }\label{dist_com}
\begin{tabular}{l l l l}\hline
Cloud           		&Distances quoted                              		&Associated            &Ref. 	  \\
Identification                		&in the literature (pc)                          		&star                       &       	  \\
\hline\hline
L 1517      		&$144^{+23, \dag}_{-17}$                       		&AB Aur                   &1     	    \\
Chamaeleon 1          &115-215, 158                                  		&HD 97048             &         2,3    \\
                    		&$150\pm15$, $175^{+27}_{-20}$		&                             &4, 1   	     \\
Lupus 3         	        &130-170, 140                                  		&HR 5999                &5,6,7 	\\
                    		&100, $208^{+46}_{-32}$, $140\pm2$    	&                       	&8,9,10     \\
NGC 7023    		&440, $429^{+159}_{-90}$, $288\pm25$	&HD 200775		&11,1     	 \\
                   		&$288\pm25$                                              	&                  		&12          \\
\hline
\end{tabular}

1-Bertout et al. (1999);2-Schwartz (1991); 3-Franco (1991); 4-Whittet et al. (1997);
5-Murphy et al. (1986); 6-Krautter (1991); 7-Hughes et al. (1993); 8-Knude \& H{\o}g (1998);
9-van den Ancker et al. (1998); 10-de Zeeuw et al. (1999); 11-Viotti (1969); 12-Strai\v{z}ys et al. 1992.\\
$^{\dag}$The Hipparcos distance to the AB Aur.
\end{table}
The central coordinates, total number of stars selected after applying all the selection  criteria ($\sigma\leq0.035$, $SNR>10$, \& ($J-K_{s}$)$\leq0.75$) and the number of stars classified as dwarfs and that were used to determine distances to the individual clouds are  listed in columns 2-5 of the Table \ref{cha1_tab1} respectively. 
\begin{table}
\centering
\caption{The central coordinates, the number of stars selected from each field after applying all the selection criteria ($\sigma\leq0.035$, $SNR>10$, \& ($J-K_{s}$)$\leq0.75$) and the stars classified as dwarfs using our method and used to determine distances to L1517, Cha I, Lup 3 and NGC 7023 are given.}\label{cha1_tab1}
\begin{tabular}{lcccr}\hline
Field &$\alpha(J2000)$&$\delta(J2000)$    &Total&Stars classified\\
Id.   &($^{\circ}$)   &($^{\circ}$)       &stars&as dwarfs \\
\hline\hline
\multicolumn{5}{l}{L1517}          \\
F1   &74.166973 & $+$30.557486&395&292 \\
F2   &73.670855 & $+$31.053637&339&260 \\
F3   &73.555237 & $+$30.138617&508&402 \\
F4   &73.080485 & $+$30.648516&299&230 \\
Total&                 &                      &1541&1184  \\\hline
\multicolumn{5}{l}{Chamaeleon I}          \\
F1   &169.399281&$-$76.719727&521  &360   \\
F2   &165.432583&$-$76.312943&494  &352   \\
F3   &167.607235&$-$77.711487&235  &161   \\
F4   &163.267247&$-$77.305313&360  &260   \\
F5   &165.209741&$-$78.675217&502  &351   \\
F6   &160.921746&$-$78.242210&550  &389   \\
Total&                  &                      &2662&1873  \\\hline
\multicolumn{5}{l}{Lupus 3}               \\
F1   &242.405769&$-$39.179062&490  &335   \\
F2   &241.008065&$-$39.471130&814  &561   \\
F3   &243.129521&$-$38.354233&487  &308   \\
Total&          &            &1791 &1204  \\\hline
\multicolumn{5}{l}{NGC 7023}   \\
F1   &315.431610&$+$68.160049&310  &230   \\
F2   &314.744713&$+$67.160805&394  &281   \\
\hline
\end{tabular}
\end{table}
\subsection{L 1517}

\begin{figure}
\resizebox{9cm}{8.5cm}{\includegraphics{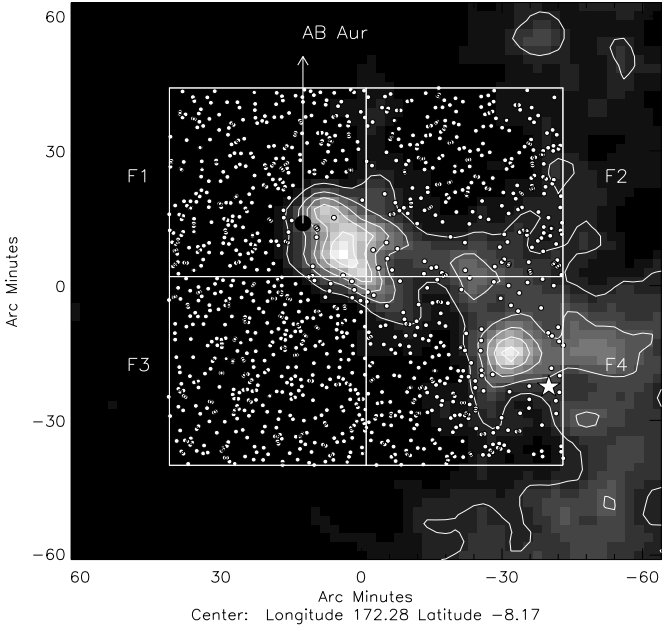}}
\caption{The $2^{\circ}\times2^{\circ}$ extinction map produced by Dobashi et al (2005) containing L1517 is shown with the fields  F1-F4, each covering $40^{\prime}\times40^{\prime}$ area, marked and labeled. The contours are drawn at 0.5, 1.0, 1.5 and 2.0 magnitude levels. The stars used for estimating distance to  L1517 are represented by filled circles. The position of the Herbig Ae star, AB Aur, is also identified and labeled. The star with high extinction ($A_{V}>1$ magnitude) at a distance closer than the estimated distance of the cloud is identified using the star symbols.}\label{Ab_img}
\end{figure}
\begin{figure}
\resizebox{9cm}{9.5cm}{\includegraphics{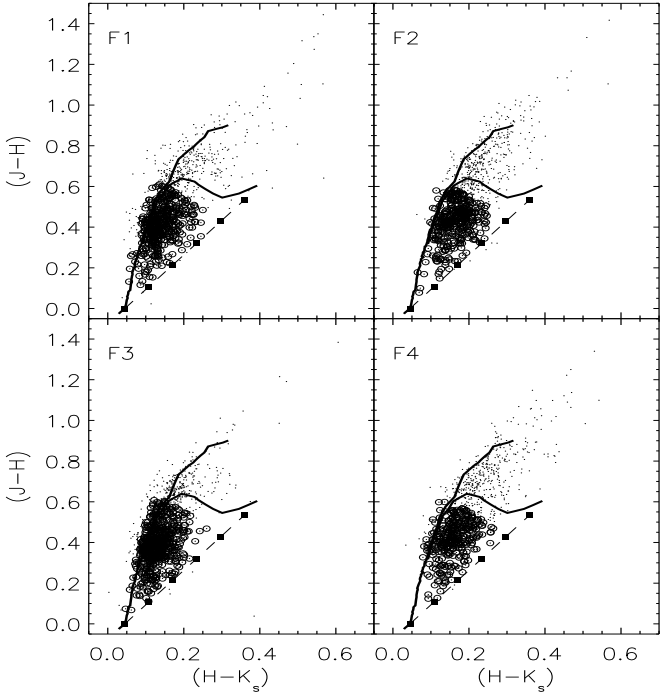}}
\caption{The NIR-CC diagrams for the stars selected from the four fields containing L1517 are shown. The dots represent all the stars in a given field with photometric errors $\leq0.035$ magnitude in $J$, $H$, \& $K_{s}$ bands and open circles represent stars with their $(J-K_{s})\leq0.75$.}\label{ab_CC}
\end{figure}
\begin{figure}
\resizebox{9cm}{7cm}{\includegraphics{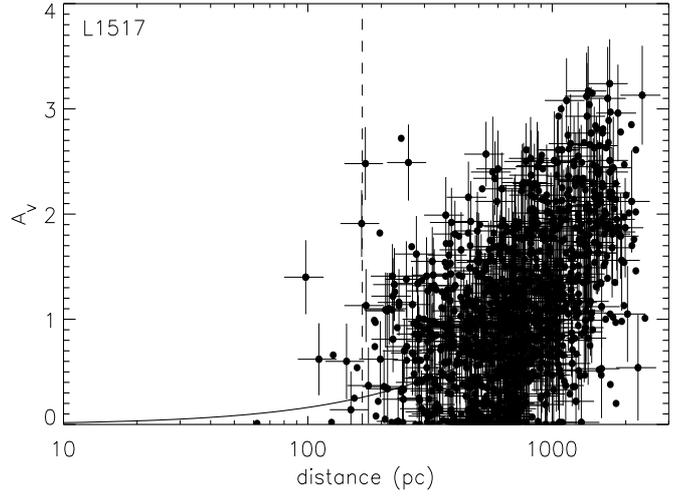}}
\caption{The $A_{V}$ vs. $d$ plot for all the stars obtained from the fields F1-F4 combined together towards L1517. The dashed vertical line is drawn at 167 pc inferred from the procedure described in the \S\ref{quant_approach}. The solid curve represents  the increase in the extinction towards the Galactic latitude of $b=-7.98^{\circ}$ as a function of distance  produced from the expressions given by Bahcall \& Soneira (1980). The error bars are not shown on all the stars for better clarity.}\label{ab_dist_all}
\end{figure}
\begin{figure}
\resizebox{9cm}{15cm}{\includegraphics{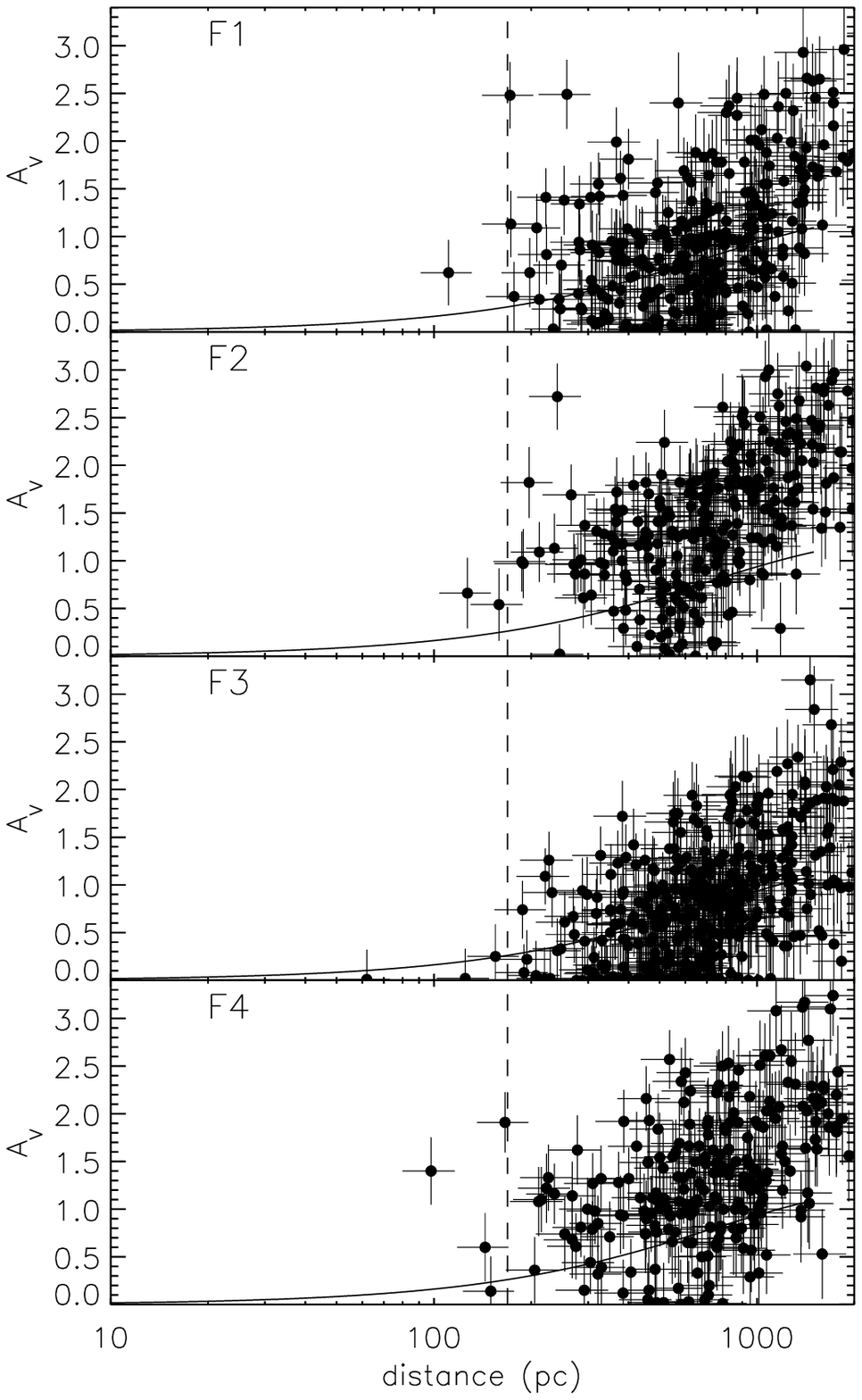}}
\caption{The $A_{V}$ vs. $d$ plots for the stars from the fields F1-F4 towards L1517. The dashed vertical line is drawn at 167 pc inferred from the procedure described in the \S\ref{quant_approach}. The solid curve represents  the increase in the extinction towards the Galactic latitude of $b=-7.98^{\circ}$ as a function of distance produced from the expressions given by Bahcall \& Soneira (1980).}\label{ab_dist}
\end{figure}

In Fig. \ref{Ab_img} we present the $2^{\circ}\times2^{\circ}$ high resolution extinction map~\footnote{The extinction map, covering the entire region in the galactic latitude range $|b|\leq40^{\circ}$ derived using the optical database ``Digitized Sky Survey I'' and the traditional star-count technique, was produced in two angular resolutions of $6^{\prime}$ and $18^{\prime}$ (Dobashi et al. 2005).} produced by Dobashi et al. (2005) of the region containing L1517. The contours are drawn at 0.5, 1.0, 1.5 and 2 magnitude levels. The Herbig Ae star, AB Aur (Herbig 1960; The et al 1994), associated with the reflection nebulosity, GN 04.52.5.02 (Magakian 2003), is identified in Fig. \ref{Ab_img}. 

Some of the stars classified as main sequence by our method could actually be giants. Such erroneous classifications  could lead to an underestimation of their distances which could create confusion with the real increase  in the extinction due to the presence of a cloud. This problem could be circumvented if we divide a large field, containing the cloud, into small sub-fields. While the rise in the extinction due to the presence of a cloud should occur almost at the same distance in all the fields, if the whole cloud is  located at the same distance, the wrongly classified stars in the sub-fields would show high extinction not at same but random  distances. We divided the region containing L1517 into four fields as shown in Fig. \ref{Ab_img}. Each field is of  $40^{\prime}\times40^{\prime}$ in area. The central coordinates,  the  number of stars selected after applying all the selection  criteria ($\sigma\leq0.035$, $SNR>10$, \& ($J-K_{s}$)$\leq0.75$) and the number of stars classified as dwarfs by our method from each field are listed in Table \ref{cha1_tab1}. In Fig. \ref{ab_CC} we show the NIR-CC diagrams for the stars selected from F1-F4.  The stars with their photometric errors $\leq0.035$ magnitude in $J$, $H$, \& $K_{s}$ bands are shown as dots and  among them the stars with $(J-K_{s})\leq0.75$ are identified by open circles. The stars classified as dwarfs using our method are identified by filled circles (in white) in Fig. \ref{Ab_img}. It can be noticed that very few sources are  selected towards the high extinction regions  due to our data selection criteria discussed in section \ref{data}.

In Fig. \ref{ab_dist_all} we show the $A_{V}$ vs. $d$ plot for the stars from all the fields combined together. The solid curve shows  
the increase in the extinction towards the Galactic latitude  $b=-7.98^{\circ}$ as a function of distance  produced using the expressions given by Bahcall \& Soneira (1980) (BS80, hereafter). Clearly, the stars with high extinction located beyond 167 pc, marked by the vertical dashed line (drawn using the procedure described in the \S\ref{quant_approach}, see Fig. \ref{4clds}), are distributed almost continuously. There exist one star with  $A_{V}>1$ magnitude at distance closer than 167 pc. The $A_{V}$ vs. $d$ plots for the stars from the  individual fields F1-F4 are shown in the Fig. \ref{ab_dist}. The stars showing high extinction at 167 pc are located in the fields, F1 and F4, which contain the major parts of the cloud. The high extinction star closer than 167 pc in the field F4 is identified using filled star symbol in the Fig. \ref{Ab_img}. Because the star is projected within the cloud boundary, it is likely to be a misclassified source located behind the cloud.  

The distances to the molecular clouds in the Taurus-Auriga complex is considered as 140 pc based on the star count, photometric distances of reflecting nebulae and reddening versus distance diagrams of the field stars towards the entire complex. But the most reliable estimation of the distance to L1517 is based on the Hipparcos parallax measurements of the two stars, AB Aur and SU Aur, found associated  with the cloud. The Hipparcos distances to these stars are $144^{+23}_{-17}$ and $152^{+63}_{-34}$ respectively. The distance of  $167\pm30$ pc to L1517 determined from our method are found to be in good agreement with the distances of AB Aur and SU Aur within the uncertainties.


\subsection{Chamaeleon 1}

\begin{figure}
\resizebox{9cm}{9cm}{\includegraphics{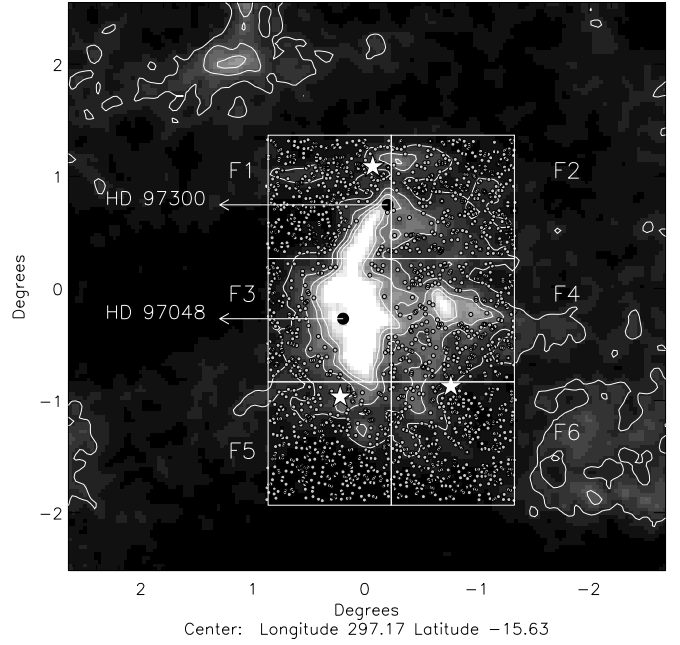}}
\caption{The $5^{\circ}\times5^{\circ}$ extinction map produced by Dobashi et al (2005) containing Cha I is shown with the contours drawn at 0.5, 1.0, 1.5 and 2.0 magnitude levels. The fields F1-F6 selected towards Cha I are marked and labeled in the figure. The stars classified  as dwarfs using our method are identified with filled circles. Each field is of $1^{\circ}\times1^{\circ}$ in area. The locations of the stars HD 97300 \& HD 97048  are identified and labeled. The stars with high extinction ($A_{V}>1$ magnitude) at distances closer than the estimated distance of the cloud is identified using the star symbols.}\label{Cha1_img}
\end{figure}
\begin{figure}
\resizebox{9cm}{7cm}{\includegraphics{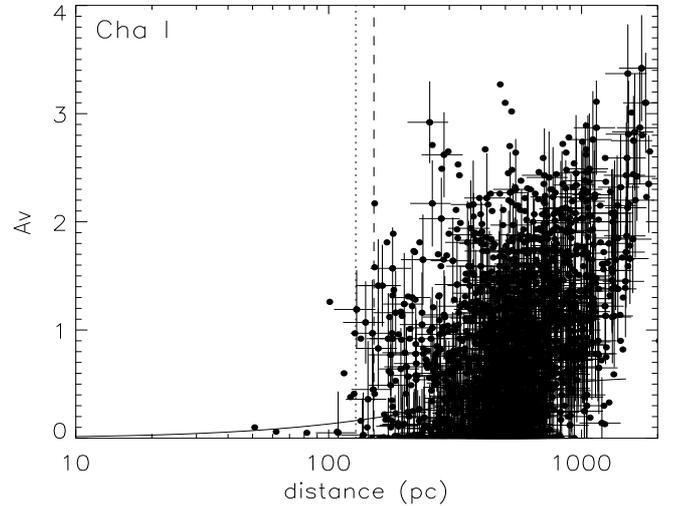}}
\caption{The $A_{V}$ vs. $d$ plot for the stars from all the fields (F1-F6) of Cha I  combined together. The vertical dashed line is drawn at 151 pc inferred from the procedure described in the \S\ref{quant_approach}. The vertical dotted line is drawn to indicate the distance of a possible foreground dust layer at $\sim128$ pc. The solid curve represents the  increase in the extinction towards the Galactic latitude of $b=-15.9^{\circ}$ as a function of distance  produced from the  expressions given by BS80. For better clarity, the error bars are not shown on all the stars.}\label{Cha1_all}
\end{figure}
\begin{figure}
\resizebox{8.5cm}{22cm}{\includegraphics{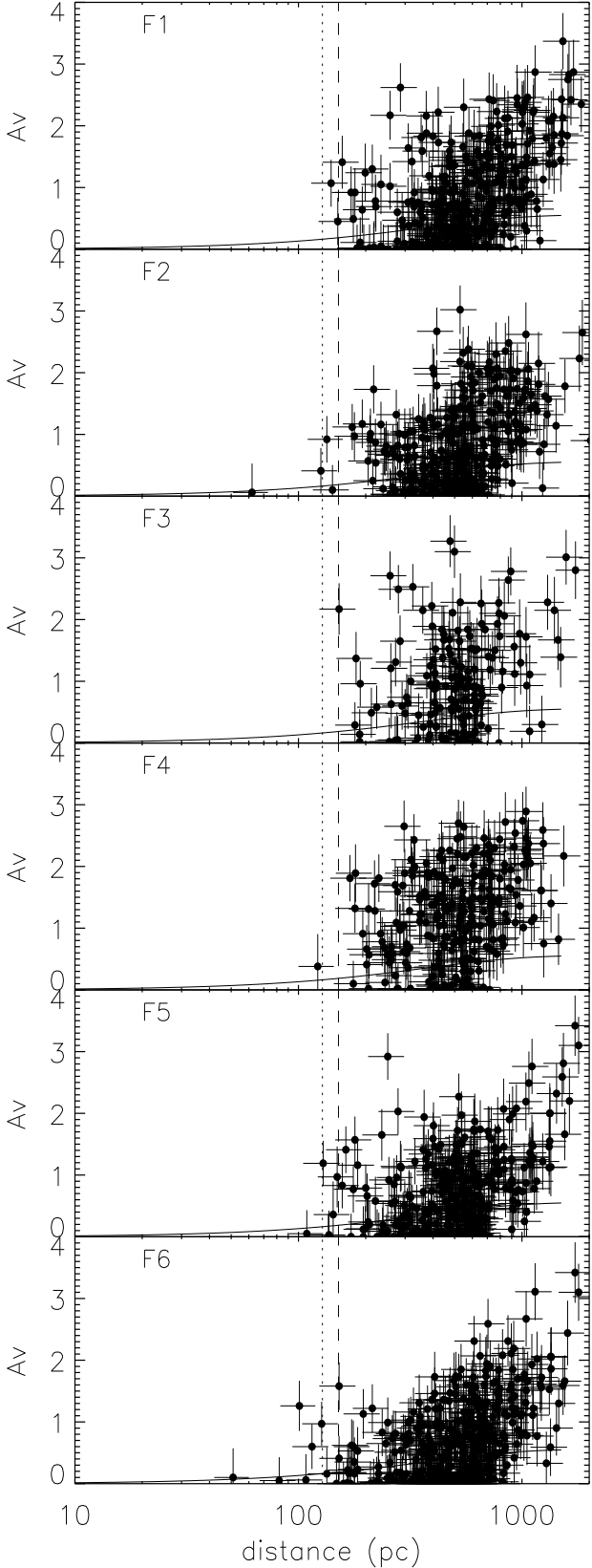}}
\caption{The $A_{V}$ vs. $d$ plot for stars in the fields F1-F6 selected towards Cha I are shown here. The dashed vertical line is drawn 151 pc inferred from the procedure described in the \S\ref{quant_approach}. The dotted vertical line is drawn a 128 pc. The solid curve represents the  increase in the extinction towards the Galactic latitude of $b=-15.9^{\circ}$ as a function of distance  produced from the expressions given by Bahcall \& Soneira (1980).}\label{Cha1_dist}
\end{figure}

The region containing the cloud Cha I is shown in the $5^{\circ}\times5^{\circ}$ extinction map produced by Dobashi et al (2005) in Fig. \ref{Cha1_img}. The contours are drawn at 0.5, 1.0, 1.5 and 2.0 magnitude levels. Since the cloud is extended over a region of  about $3^{\circ}\times2^{\circ}$ area,  we divided the whole region into six sub-fields of $1^{\circ}\times1^{\circ}$ each as identified  in Fig. \ref{Cha1_img}. The central coordinates,  total number of stars selected after applying all the selection  criteria ($\sigma\leq0.035$, $SNR>10$, \& ($J-K_{s}$)$\leq0.75$) and the number of stars classified as dwarfs using our method in each field are listed in Table \ref{cha1_tab1}. Stars classified as dwarfs are identified in Fig. \ref{Cha1_img} using filled circles. Here too we find very few sources towards the high extinction regions. 

In Fig. \ref{Cha1_all}, we show the $A_{V}$ vs. $d$ plot for stars from all the fields combined together. The solid curve represents   the increase in the extinction towards the Galactic latitude of $b=-15.9^{\circ}$ as a function of distance  produced using the expressions given by BS80. Using the procedure described in the \S\ref{quant_approach}, we found a drop in the extinction at 151 pc (marked by the vertical dashed line in the  Fig. \ref{Cha1_all}) as shown in the Fig. \ref{4clds}. But  the drop in the extinction at 151 pc was not as conspicuous as was observed towards L1517. Three stars with $A_{V}\geq1$ were found to be located closer than 151 pc (identified in Fig. \ref{Cha1_img} using filled star symbols). We noticed  another drop in the extinction at 128 pc (see Fig. \ref{4clds}). In Fig. \ref{Cha1_dist}, we show $A_{V}$ vs. $d$ plot for stars in the individual fields. The dotted line is drawn at 128 pc. The stars with extinction significantly above the values expected from the BS80 model seem to be present in all the fields at or beyond 151 pc unlike the distribution of the stars closer than 151 pc implying that the foreground dust layer at 128 pc is very patchy in distribution. 
 
Earlier attempts to determine distances to Cha I were made by Whittet et al. (1987), Franco (1991), Schwartz (1991) and Whittet et al. (1997). Schwartz (1991) reported a value of 115-215 pc, Franco (1991) assigned 158 pc while Whittet et al. (1997) deduced a most probable distance of $150\pm15$ pc based on the reddening vs. distance plot for stars towards Cha I. Independent distance estimates based on parallax measurements made by the Hipparcos satellite are available for two stars, HD 97048 and HD 97300  (van den Ancker et al. 1998; Bertout et al. 1999), which illuminate prominent reflection nebulae, Ced 111 and Ced 112 respectively. The parallax distances of HD 97048 and HD 97300 were estimated to be $175^{+27}_{-20}$ and $188^{+44}_{-30}$ respectively (Bertout et al. 1999).  But the spectrophotometric distances to these stars were estimated to be $180\pm20$ and $152\pm18$ respectively (Whittet et al. 1997). These stars are located in   F3 and F1 as identified in Fig. \ref{Cha1_img}. Based on $uvby\beta$ photometry of 1017 stars, Corradi et al. (1997) showed the presence of a dust layer which is a part of a large scale distribution extended over a large area covering Chamaeleon, Musca and Coalsack at a distance of $\sim150$ pc. In a subsequent work (Corradi et al. 2004) they showed the presence of gas as close as 60 pc from the Sun using interstellar Na I D absorption lines towards 63 B-type stars. The dust layer found in Fig. \ref{Cha1_all} and \ref{Cha1_dist} at 128 pc could be a part of the forground dust layer shown by Corradi et al. (1997, 2004). The onset of extinction in $A_{V}$ vs. $d$ plots for both F1 and F3 are found to occur consistently at $\sim151$ pc (Fig. \ref{Cha1_dist}).  Based on our method, the  present determination of $151\pm28$ pc to Cha 1 is found to be in good agreement with the most probable distance estimated for the cloud by Whittet et al. (1997) {\bf and Corradi et al. (1997, 2004)}. 

\subsection{Lupus 3}

\begin{figure}
\resizebox{9cm}{8.5cm}{\includegraphics{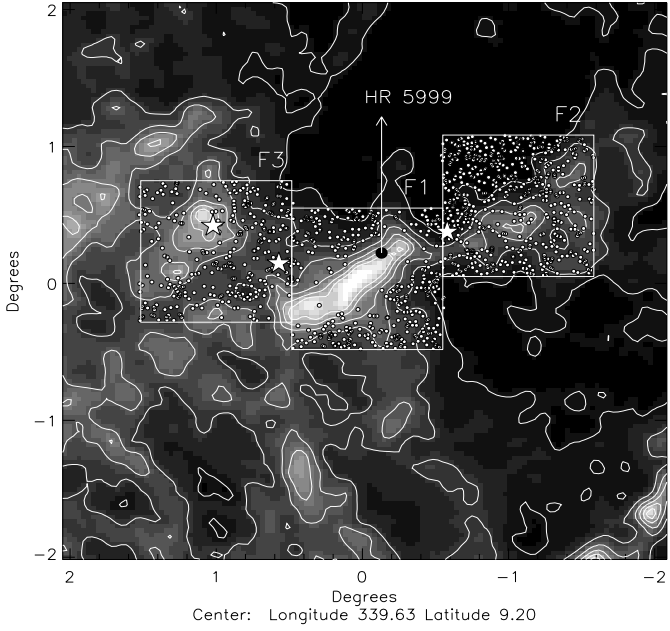}}
\caption{The $4^{\circ}\times4^{\circ}$ extinction map produced by Dobashi et al (2005) containing Lup 3 is shown with the contours drawn at 0.5, 1.0 , 1.5, 2.0, 2.5 and 3.0 magnitudes.  The three fields F1-F3 each of $1^{\circ}\times1^{\circ}$ area containing the cloud and the Herbig Ae/Be star, HR 5999, are identified and labeled in the figure. The stars with high extinction ($A_{V}>1$ magnitude) at distances closer than the estimated distance of the cloud is identified using the star symbols.}\label{Lup_img}
\end{figure}
\begin{figure}
\resizebox{9cm}{7cm}{\includegraphics{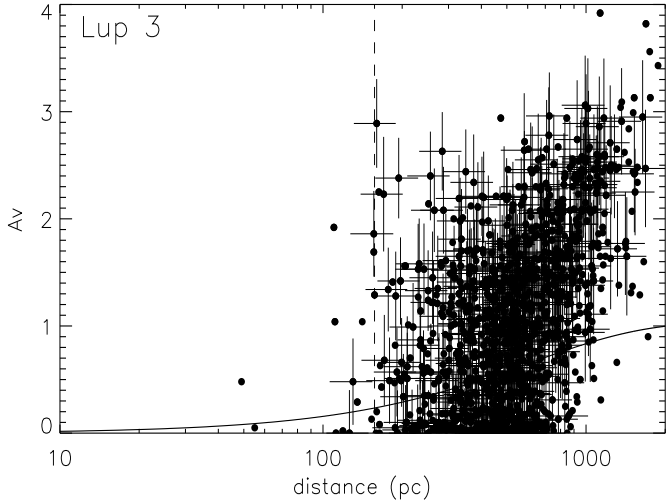}}
\caption{The $A_{V}$ vs. $d$ plot for the stars from all the fields F1-F3 of Lup 3 combined together. The dashed line is drawn at 157 pc
inferred from the procedure described in the \S\ref{quant_approach}. The error bars are not shown on all the stars for better clarity.}\label{Lup_all}
\end{figure}
\begin{figure}
\resizebox{9cm}{15cm}{\includegraphics{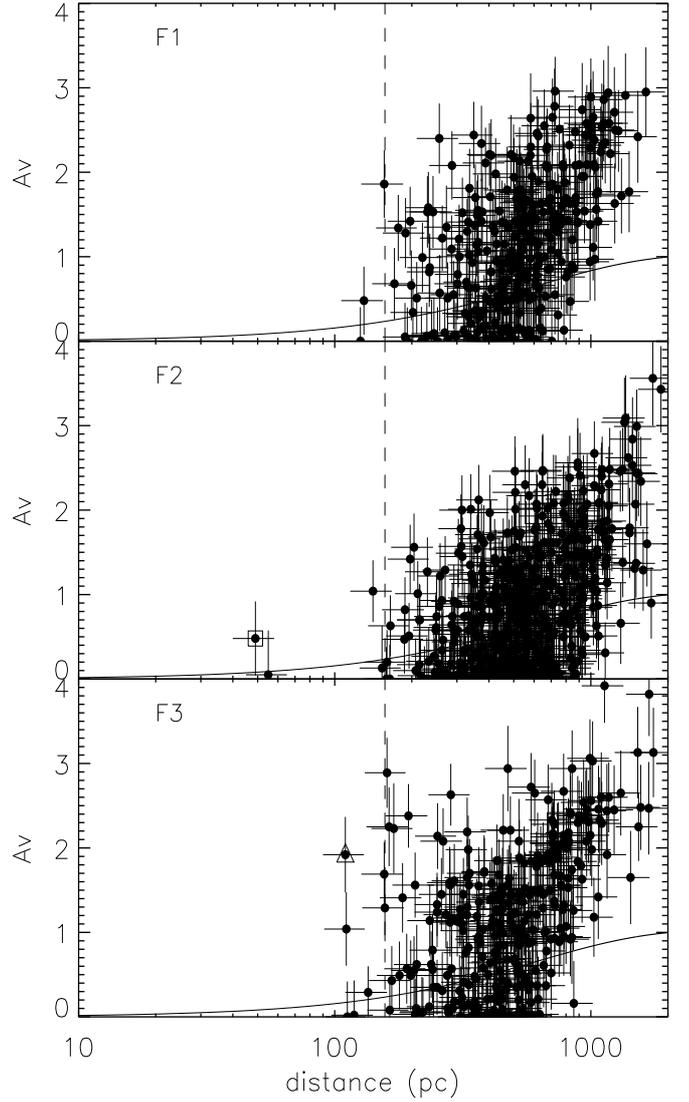}}
\caption{The $A_{V}$ vs. $d$ plots for stars from the individual fields F1- F3 of Lup 3. The dashed line is drawn at 157 pc
inferred from the procedure described in the \S\ref{quant_approach}. The dot-dashed line shows the increase in extinction towards the Galactic latitude of $b=+9.4^{\circ}$ as a function of distance  estimated using expressions given by BS80. The source marked with an open triangle is identified with HD 143261, a giant (K1/K2III).}\label{Lup_dist}
\end{figure}

The region containing the Lup 3 cloud is shown in the $4^{\circ}\times4^{\circ}$  extinction map produced by Dobashi et al (2005) in Fig. \ref{Lup_img}. The contours are drawn at 0.5 to 3.0 with an interval of 0.5 magnitude. The three fields containing the cloud are identified in square boxes. The field, F1 contains Herbig Ae star, HR 5999 (The et al 1994). The central coordinates, total number of stars selected after applying all the selection criteria ($\sigma\leq0.035$, $SNR>10$, \& ($J-K_{s}$)$\leq0.75$) and the number of stars classified as dwarfs (identified in Fig. \ref{Lup_img} with filled circles) are listed in Table \ref{cha1_tab1}. As seen in the cases of L1517 and Cha I, towards Lupus 3 too, we found very few stars towards the dense parts of the cloud. 

In Fig. \ref{Lup_all}, we show the $A_{V}$ vs. $d$ plot for the stars from all the three fields combined. The solid curve represents  the increase in the extinction towards the Galactic latitude of $b=+9.4^{\circ}$  as a function of distance produced using the expressions given by BS80. We found a sharp drop in the extinction at 157 pc, as shown in Fig. \ref{4clds}, estimated using the procedure described in the \S\ref{quant_approach} (marked with the vertical dashed line in the Fig. \ref{Lup_all}). There are  three stars with $A_{V}\geq1$ at distances closer than 157 pc in Fig. \ref{Lup_all}. In Fig. \ref{Lup_dist}, we show the  $A_{V}$ vs. $d$ plot for the individual fields towards the Lup 3.  The source located at 49 pc and $A_{V}=0.5$ in F2 is identified with HD 143261, a giant, classified as K1/K2III (Simbad database). Of the two stars with $A_{V}\geq1$ in F3, the one located at 110 pc and $A_{V}=1.9$ is identified with HD 145355 classified as A1III (Simbad database).

Most of the earlier studies have determined or assumed distances in the range 130-170 pc either by considering Lupus to be associated with Scorpius-Centaurus association  (Murphy et al. 1986; Krautter 1991) or from spectroscopic parallaxes of early-type stars thought to be members of Scorpius-Centaurus association (Hughes et al. 1993). But Knude \& H{\o}g (1998) based on extinction of stars located in the foreground and the background of Lupus complex with their distances estimated from Hipparcos parallaxes, determined a  closer distance of 100 pc to Lup 3. However, the Hipparcos parallax distance of HR 5999, illuminating the nebula GN 16.05.2 (Magakian 2003),  is estimated to be $208^{+46}_{-32}$ pc (van den Ancker et al. 1998) somewhat farther than most of the distance estimates. The best distance estimation currently available for the Centaurus-Lupus subgroup of the Scorpius-Centaurus association is $140\pm2$ pc derived by de Zeeuw et al. (1999) based on Hipparcos parallaxes. Using our method we determined a distance of $157\pm28$ pc to Lup 3 cloud.

\subsection{NGC 7023}
\begin{figure}
\resizebox{9cm}{8.5cm}{\includegraphics{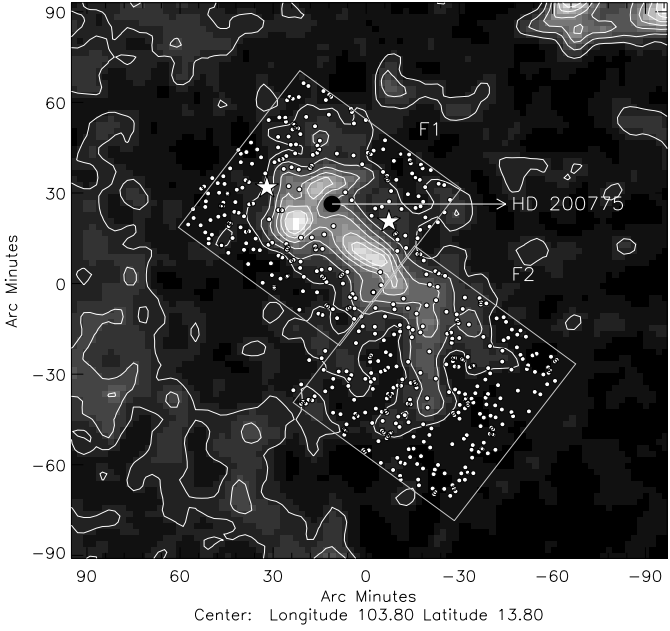}}
\caption{The $3^{\circ}\times3^{\circ}$ extinction map of the region containing NGC 7023 is shown with the fields F1 and F2 ( of $1^{\circ}\times1^{\circ}$ each) identified. The contours are drawn at 0.6, 1.0, 1.5, 2.0, 2.5 and 3.0 magnitude. The arrow shows the location of the Herbig Ae/Be star HD 200775. The stars (only from F1) with high extinction ($A_{V}>1$ magnitude) at distances closer than the estimated distance of the cloud is identified using the star symbols.}\label{NGC70_img}
\end{figure}
\begin{figure}
\resizebox{9cm}{12cm}{\includegraphics{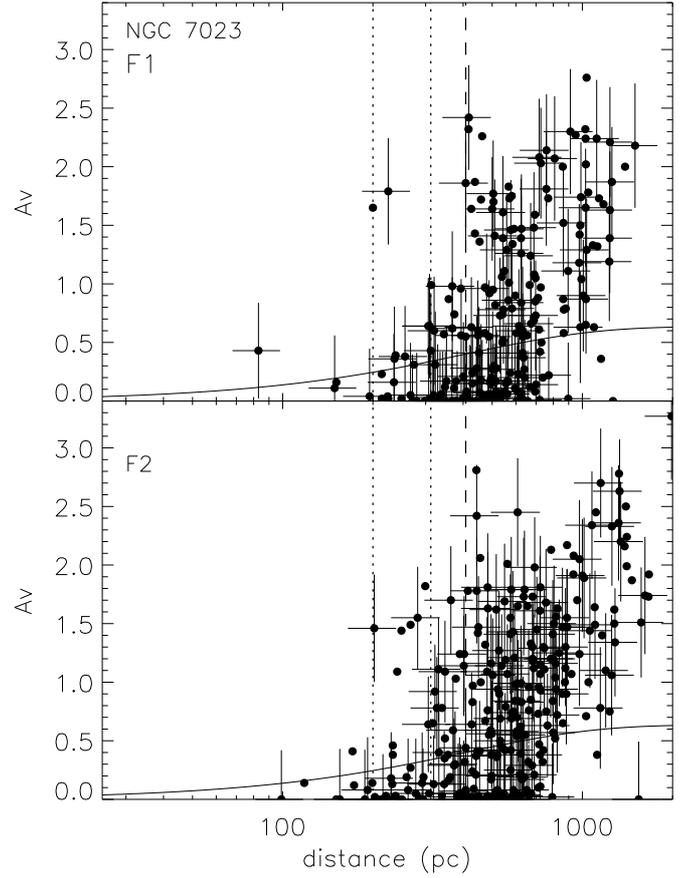}}
\caption{The $A_{V}$ vs. $d$ plots for the stars in F1 and F2 towards NGC 7023.  The dashed lines is drawn at 408 pc inferred from the
the procedure discussed in the \S\ref{quant_approach}. The dotted lines indicate the locations of two additional dust layers possibly present at $\approx200$ pc  and $\approx310$ pc. The solid curve represents the  increase in extinction towards the Galactic latitude of $b=+14.1^{\circ}$ as a function of distance  estimated using expressions given by BS80. The error bars are not shown on all the stars for better clarity.}\label{NGC70_dist}
\end{figure}

The reflection nebula NGC 7023 is illuminated by the Herbig Be star HD 200775 (The et al. 1994). The  $3^{\circ}\times3^{\circ}$ extinction map of the region containing NGC 7023 is shown in  Fig. \ref{NGC70_img}. The contours are drawn at 0.6, 1.0, 1.5, 2.0, 2.5 and 3.0 magnitude. Two fields of $1^{\circ}\times1^{\circ}$ were chosen towards the region. The central  coordinates, total number of stars selected after applying all the selection  criteria ($\sigma\leq0.035$, $SNR>10$, \& ($J-K_{s}$)$\leq0.75$) and the number of stars classified as dwarfs are listed in the Table \ref{cha1_tab1}.  These stars are identified with filled circles in Fig. \ref{NGC70_img}.

In Fig. \ref{NGC70_dist}, we present the $A_{V}$ vs. $d$ plot for stars from the fields F1 and F2. The solid curve shows the increase in the extinction towards the Galactic latitude  of $b=+14.1^{\circ}$ as a function of distance produced using the expressions given by BS80. Clearly, a drop in the extinction is noticeable at two distances for  F1, at $\approx310$ pc and at $\approx410$ pc. In F2, though there exist a step like appearance again at distances $\approx310$ and $\approx410$ pc, a number of stars show relatively high extinction at  much closer distance of $\approx200$ pc, a distance where an increase in the  $A_{V}$ of $\geq1.5$ magnitude is seen in F1 also but for only two stars. From  the $A_{V}$ vs. $d$ plot for stars in F1 and F2, we infer the presence of at least three layers of dust grains along the line of sight towards NGC 7023. While the dust components at  $\approx310$ pc and $\approx410$ pc are dominant towards F1, the components at $\approx200$ pc and  $\approx310$ pc are dominant towards F2. Using the procedure discussed in the \S\ref{quant_approach}, we found a significant drop  in the mean value of  $A_{V}$ at 408 pc (the distance marked by the vertical dashed line). We used the stars from F1 alone as the situation is much complex towards F2.

The most widely quoted distances to LDN 1167/1174 or NGC 7023 and LDN 1147/1158  (clouds located $\sim2^{\circ}$ west of NGC 7023) groups are $288\pm25$ and $325\pm13$ pc respectively (Strai\v{z}ys et al. 1992). They used the Vilnius photometric system to classify the stars and also to get the interstellar reddening. They obtained a distance of $325\pm13$ pc to LDN 1147/1158 by taking an average of 10 stars showing $A_{V}\geq0.45$ which were distributed in the range 240-380 pc. For LDN 1167/1174 or NGC 7023 group, they assigned a  distance of $288\pm25$ pc again by taking the average distance of 4 considerably reddened stars. They preferred a distance of 275 pc to HD 200775 by assuming it to be a B3Ve star.

The Hipparcos parallax distance to HD 200775 is estimated to be $429^{+156}_{-90}$ pc (van den Ancker at al. 1998; Bertout et al. 1999). The sharp rise in the extinction in Fig. \ref{NGC70_dist} at 408 pc could be due to the association of the cloud with HD 200775 and the sharp rise in $A_{V}$ at 305 pc could be due to a foreground dust component. Kun (1998) showed the presence of dust components at three characteristic distances: 200, 300, and 450 pc, based on cumulative distribution of field star distance moduli in Wolf diagrams. We found an additional component of dust at 170 pc towards the western parts of NGC 7023 especially towards LDN 1171 and LDN 1147/1158 group. The results will be presented in a forthcoming paper. Using a total of 230 stars, classified as dwarfs, we determined a  distance of $408\pm76$ pc to the NGC 7023 cloud.

\section{Discussion}\label{discuss}
\begin{figure}
\resizebox{9cm}{15cm}{\includegraphics{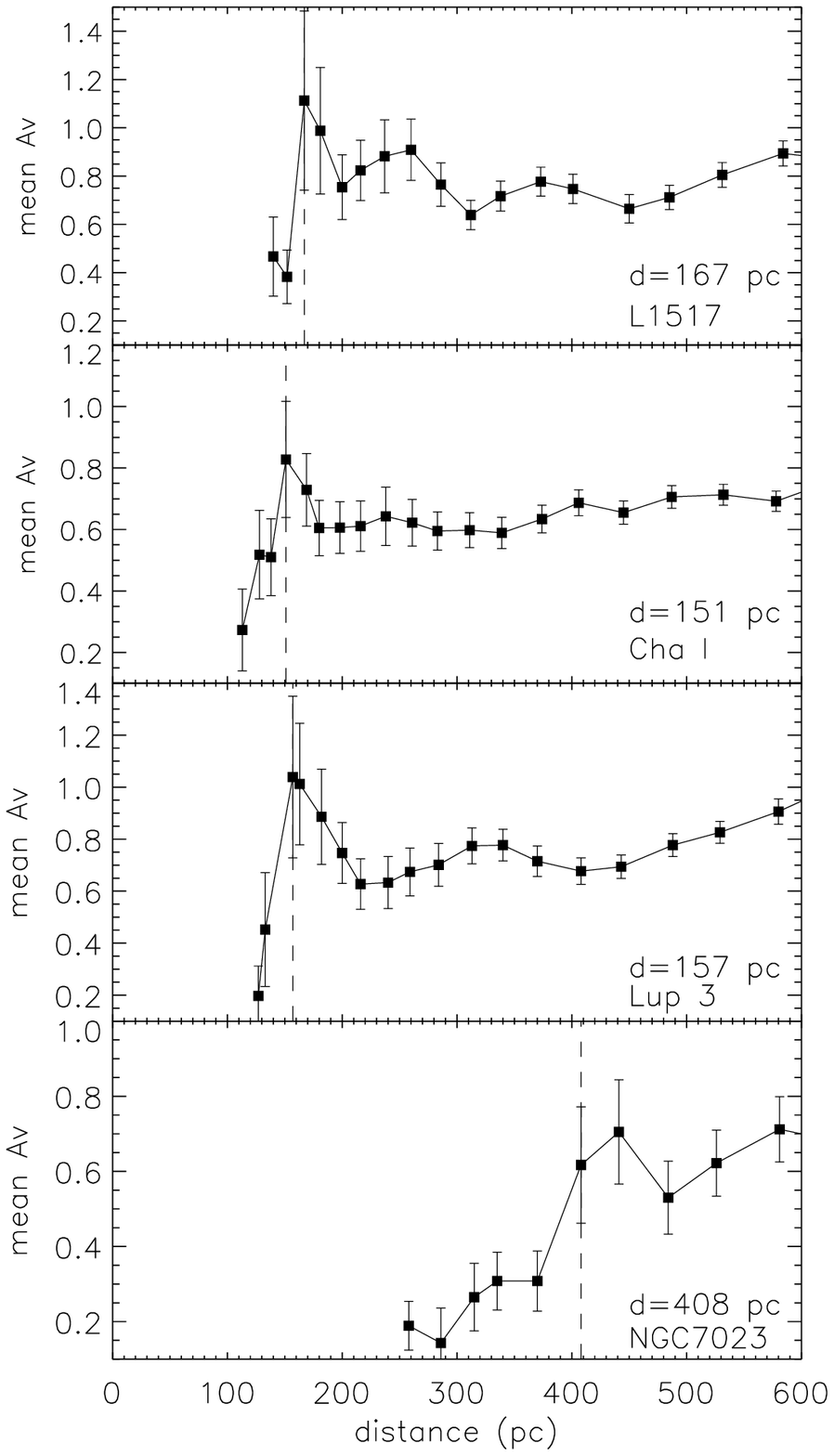}}
\caption{The mean values of $A_{V}$ vs. the mean values of distance plot for L1517, Cha I, Lupus 3 and NGC 7023 produced using the the procedure discussed in the \S\ref{quant_approach} to determine distances to the clouds. The error bars on the mean $A_{V}$ values were calculated using the expression, $standard~deviation/\sqrt[]{N}$, where $N$ is the number of stars in each bin.}\label{4clds}
\end{figure}

The current determination of distances to L1517, Cha I, Lupus 3 and NGC 7023 is based on the NIR photometry of relatively large number of stars, 1184, 1873, 1204 and 230, respectively  (Table \ref{cha1_tab1}), projected onto them. Distances estimated using our method to these clouds are found to be in good agreement with most of the previous estimations within the error (see Table \ref{dist_com}). 

Based on our method and the NIR data from the 2MASS, one could determine distances to any cloud located within  $\sim500$ pc from the Sun (depending upon the Galactic latitude of the clouds).  However, the complex nature of the interstellar medium is clearly evident  in the $A_{V}$ vs. $d$ plots of Cha 1 and NGC 7023, as shown in Figures \ref{Cha1_all} \& \ref{NGC70_dist}, respectively. There exist, clearly, at least two layers of dust components in the direction towards these two regions. If the first layer of dust component happens to be denser but uniform in distribution, it would reduce the probability of finding the second dust layer. However, if the first layer is denser but patchy in  distribution, we get sources with high extinction at shorter distances before we could notice a wall of stars with high extinction due to the second layer of dust component. In that case, differentiating  stars with spurious spectral classification from the  actual rise in extinction due to the presence of dust components would become  difficult. Dividing a larger field into sub-fields could help us decipher the $A_{V}$ vs. $d$ plots  of a cloud. It is indeed noted that the presence of dominant dust components is revealed in $A_{V}$ vs. $d$ plots for different fields towards Cha 1 (Fig. \ref{Cha1_all}) and NGC 7023 (Fig. \ref{NGC70_dist}) where the effects are most conspicuous. 

The size of the sub-fields required towards a cloud should be decided on the basis of its galactic location. We noticed that in a given field, our method consistently assigns $\approx70\%$ of the total selected stars, after applying all the selection criteria ($\sigma\leq0.035$, $SNR>10$, \& ($J-K_{s}$)$\leq0.75$), as dwarfs.  Our experience from the application of the method to the four clouds in the previous sections showed that we require at least 100-200 dwarfs to discern the presence of a sudden rise (or drop) in the  extinction in $A_{V}$ vs. $d$ plot of a field. The number of stars decreases with the Galactic  latitude, thus, requiring a larger field as we go towards relatively high galactic latitudes.  At higher galactic latitudes, however, the confusion due to multiple dust components in the line of sight  would be reduced as the majority of the clouds tend to occupy regions closer to the Galactic plane.

For small and isolated clouds, it would be difficult to divide the field containing the  cloud into sub-fields with sufficient number of stars projected on to them and therefore the determination of distances  made correspondingly hard. In such cases, clouds which are spatially closer and show similar radial velocities are to be chosen. Clouds located in the close proximity with similar radial velocities are believed to be at similar distances.

\section{Conclusions}\label{conclu}
We present a method to determine distances to molecular clouds using the prodigious  NIR data provided by the 2MASS. The method involves the following steps:
\begin{enumerate}
\item Extract $J$, $H$, $K_{s}$ magnitudes with photometric uncertainty $\leq0.035$ and $SNR>10$ of stars projected onto the fields containing the cloud from the 2MASS database.

\item Select stars with their ($J-K_{s}$)$\leq0.75$ to eliminate M-type stars from the analysis as unreddened M-type stars located across the reddening vectors of A0-K7 dwarfs make it difficult to differentiate the reddened A0-K7 dwarfs from the unreddened M-type stars.

\item A set of dereddened colours for each star is produced from their observed colours by using a range of trial values of $A_{V}$ (0-10 mag) and the Rieke \& Lebofsky (1985) extinction law in equations \ref{eq2} and \ref{eq3}.

\item Computed sets of dereddened colours of a star are then compared with the intrinsic colours of the normal main sequence stars. The intrinsic colours of the main sequence stars in the spectral range  A0-K7 are produced from the 2MASS data of stars with known spectral types and those located within 100 pc, calculated from the Hipparcos parallaxes. The best match giving a minimum value of $\chi^{2}$ as defined in the equation \ref{eq_chi1} then yields the spectral type and $A_{V}$ corresponding to that intrinsic colour. 

\item Once the spectral types and $A_{V}$ values of the stars are known, their distances are estimated using the distance equation \ref{eq1}. Only those stars which are classified as dwarfs are considered for the determination of distances as the absolute magnitudes of giants are highly uncertain.
\end{enumerate}

We applied the method first to main sequence stars which are located at distances $\leq100$ pc  (from Hipparcos parallaxes) and have their spectral types known. We showed that our method can determine the spectral types of the stars with a $1\sigma$ uncertainty of about four subclasses which propagates into the uncertainty in the distances of the stars by $\leq20\%$. By applying the method   to four clouds, L1517, Chamaeleon I, Lupus 3 and NGC 7023 which harbour nebulous stars with the Hipparcos parallax distances known, we determined a distance of $167\pm30$, $151\pm28$,  $157\pm29$ and $408\pm76$ pc, respectively.  These are found to be in good agreement with the most accurate distances available for them in the literature.  We found additional components of dust layers at $\approx100$ pc towards Chamaeleon I and at $\approx200$ and $\approx310$ pc towards NGC 7023. In the future, we will extend the application of this method to many more molecular clouds to determine their distances using the 2MASS NIR photometry. 

\section*{Acknowledgments}
We thank the referee, Dr. Franco, G. A. P, for his useful comments on the work. This publication makes use of data products from the Two Micron All Sky Survey, which is a joint project of the University of Massachusetts and the Infrared Processing and Analysis Center/California Institute of Technology, funded by the National Aeronautics and Space Administration and the National Science Foundation. This research has also made use of the SIMBAD database, operated at CDS, Strasbourg, France.   S. D. is supported by the MAGNET project of the ANR (France).
\section*{REFERENCES}

Alves, Jo\~{a}o., Lada, Charles J., Lada, Elizabeth A., Kenyon, Scott J., Phelps, Randy., 1998, ApJ, 506, 292\\
Alves, F. O., Franco, G. A. P., 2007, A\&A, 470, 597\\
Bahcall  J. N., Soneira  R. M., 1980, ApJS, 44, 73\\
Bertout, C., Robichon, N., Arenou, F., 1999,A\&A, 352, 574\\
Bessell, M. S., Brett, J. M., 1988, PASP, 100, 1134\\
Bok  B. J., Bok  P. F., 1941, The Milky Way. Harvard Univ. Press,  Cambridge  , MA \\
Cambr\'{e}sy, L., Beichman, C. A., Jarrett, T. H., Cutri, R. M., 2002, AJ, 123, 2559\\
Carpenter  J. M., 2001, AJ, 121, 2851\\
Clemens, Dan P., Yun, Joao Lin., Heyer, Mark H., 1991, ApJS, 75, 877\\
Corradi, W. J. B., Franco, G. A. P., Knude, J., 2004, MNRAS, 347, 1065\\
Corradi, W. J. B., Franco, G. A. P., Knude, J., 1997, A\&A, 326, 1215\\
Cox, Arthur N., Allen's astrophysical quantities, 4th ed. Publisher: New York: AIP Press; Springer, 2000. Edited by Arthur N. Cox. ISBN: 0387987460\\
Cutri  R. M.  et al., 2000, 2MASS All-Sky Catalog of Point Sources. NASA/IPAC Infrared Science Archive.\\
Dobashi, Kazuhito., Uehara, Hayato., Kandori, Ryo., et al., 2005, PASJ, 57S\\
Dutra, C. M., Santiago, B. X., Bica, E. L. D., Barbuy, B., 2003, MNRAS, 338, 253\\
de Zeeuw, P. T., Hoogerwerf, R., de Bruijne, J. H. J., Brown, A. G. A., Blaauw, A., 1999, AJ, 117, 354\\
Franco, G. A. P., 1991, A\&A, 251, 581\\
Franco, G. A. P., 2002, MNRAS, 331, 474\\
Gould, Andrew., 2003, astro.ph.10577\\
Herbig, George H., 1960, ApJS, 4, 337\\
Hilton, J., Lahulla, J. F., 1995, A\&AS, 113, 325\\
Hobbs, L. M., Blitz, L., Magnani, L., 1986, ApJ, 306L, 109\\
Hughes, Joanne, Hartigan, Patrick, Clampitt, Lori, 1993, AJ, 105, 571\\
Indebetouw, R., Mathis, J. S., Babler, B. L., et al., 2005, ApJ, 619, 931\\
Itoh, Yoichi, Tamura, Motohide, Gatley, Ian., 1996, ApJ, 465L, 129\\
Kandori, Ryo., Dobashi, Kazuhito., Uehara, Hayato., Sato, Fumio., Yanagisawa, Kenshi., 2003, AJ, 126, 1888\\
Kauffmann, J., Bertoldi, F., Bourke, T. L., Evans, N. J., II, Lee, C. W., 2008, A\&A, 487, 993\\
Kharchenko, N. V., Kinematika i Fizika Nebesnykh Tel, vol. 17, no. 5, p. 409-423\\
Kleinmann, S. G., Lysaght, M. G., Pughe, W. L., et al. 1994, Ap\&SS, 217, 11\\
Knude, J., H{\o}g, E., 1998, A\&A, 338, 897\\
Koornneef, J.,  1983, A\&A, 128, 84\\
Krautter, J., Low Mass Star Formation in Southern Molecular Clouds, ESO Scientific Report. Edited by Bo Reipurth. Garching: European Southern Observatory (ESO), 1992., p.127.\\
Kun, M., 1998, ApJS, 115, 59\\
Loinard, Laurent, Torres, Rosa M., Mioduszewski, Amy J., Rodr\'{i}guez, Luis F., Gonz\'{a}lez-L\'{o}pezlira, Rosa A., Lachaume, R\'{e}gis, V\'{a}zquez, Virgilio, Gonz\'{a}lez, Erandy., 2007, ApJ, 671, 546\\
Loinard, Laurent, Torres, Rosa M., Mioduszewski, Amy J., Rodr\'{i}guez, Luis F., 2008, ApJ, 675L, 29\\
Lombardi, M., Alves, J., Lada, C. J., 2006, A\&A, 454, 781\\
Lombardi, M., Lada, C. J., Alves, J., 2008, A\&A, 480, 785\\
Magakian  T. Y., 2003, A\&A, 399, 141 \\
Maheswar  G., Manoj  P., Bhatt  H. C., 2004, MNRAS, 355, 1272\\
Maheswar, G., Bhatt, H. C., 2006, MNRAS, 369, 1822\\
Ma\'{i}z-Apell\'{a}niz, Jes\'{u}s., 2004, PASP, 116, 859\\
Mathis, John S., 1990, ARA\&A, 28, 37\\
Megier, A., Strobel, A., Bondar, A., Musaev, F. A., Han, Inwoo, KreŁowski, J., Galazutdinov, G. A., 2005, ApJ, 634, 451\\
Meyer, Michael R., Calvet, Nuria., Hillenbrand, Lynne A.,  1997, AJ, 114, 288\\
Murphy, D. C., Cohen, R., May, J., 1986, A\&A, 167, 234\\
Nielsen, A. S., Jønch-Sørensen, H., Knude, J., 2000, A\&A, 358, 1077\\
Nishiyama, Shogo., Tamura, Motohide., Hatano, Hirofumi., et al.,  2009, ApJ, 696, 1407\\
Peterson, Dawn E., Clemens, Dan P., 1998, AJ, 116, 881\\
Rieke, G. H., Lebofsky, M. J.,  1985, ApJ, 288, 618\\
Schwartz, R. D., 1991, in \textit{Low mass star formation in southern molecular clouds}, ed. B. Reipurth, ESO Scientific Report no. 11, p. 93\\
Strai\v{z}ys, V., Wisniewski, W. Z., Lebofsky, M. J., 1982, Ap\&SS, 85, 271\\
Strai\v{z}ys, V., 1991, ppag., proc., 341\\
Strai\v{z}ys, V., Cernis, K., Kazlauskas, A., Meistas, E., 1992 BaltA, 1, 149\\
The, P. S., de Winter, D., Perez, M. R., 1994, A\&AS, 104, 315\\
Torres, Rosa M., Loinard, Laurent., Mioduszewski, Amy J., Rodríguez, Luis F.,  2007, ApJ, 671, 1813\\
van den Ancker, M. E., de Winter, D., Tjin A Djie, H. R. E., 1998, A\&A, 330, 145\\
Viotti, R.,  1969, MmSAI, 40, 75\\
Whittet, D. C. B., Kirrane, T. M., Kilkenny, D., Oates, A. P., Watson, F. G., King, D. J., 1987, MNRAS, 224, 497\\
Whittet, D. C. B. 1992, Dust in the Galactic Environment (Bristol: IOP)\\
Whittet, D. C. B., Prusti, T., Franco, G. A. P., Gerakines, P. A., Kilkenny, D., Larson, K. A., Wesselius, P. R., 1997, A\&A, 327, 1194\\
Wolf  M., 1923, Astron. Nachr., 219, 109\\
Yun, Joao Lin., Clemens, Dan P., 1990, ApJ, 365L, 73\\

\end{document}